# Analysis of Radiation Effect on the Threshold Voltage of Flash Memory Device

Nahid M. Hossain[1], Jitendra Koppu[1], Masud H Chowdhury[1]

[1]Computer Science and Electrical Engineering, University of Missouri – Kansas City, Kansas City, MO 64110, USA

Email: mnhtyd@mail.umkc.edu, jkvxf@mail.umkc.edu and masud@ieee.org

Abstract- Flash memory experiences adverse effects due to radiation. These effects can be raised in terms of doping, feature size, supply voltages, layout, shielding. The the operating point shift of the device forced to enter the logically-undefined region and cause upset and data errors under radiation exposure. In this letter, the threshold voltage shift of the floating gate transistor (FGT) is analyzed by a mathematical model.

## I. INTRODUCTION

As the device size is shrinking down to nanometer range, the impacts of radiations on circuit and device performance and reliability would become more prominent. In the recent time study of radiation hardness of micro & nano-electronic devices for extreme conditions are gaining wide spread attention. The radiation effect on the floating gate transistor (FGT) used in flash memory leads to charge loss from the programmed floating gate (FG). Due to the imposture to certain type of radiation, an extra electron/hole pair can be generated in the device. For example, a minimum of 10-Kev X-rays exposure would initiate this process. In the FGT, the radiation effect can be neglected if the oxide thickness is 40-47.5 nm [1]. But it is no longer negligible because the oxide thickness is going down to 10 nm in scaled FGTs. Radiation induced charge in the oxide depends on several physical mechanisms i.e. electron-hole pair generation and election-hole recombination. The electron-hole recombination depends on the applied electrical field and linear energy transfer. Even the effect changes over time, i.e. the change of threshold voltage ($\Delta V_{TH}$) varies over the $10^{-6} \sim 10^8$ Sec time scale. When $V_G>0V$, holes are trapped into the oxide due to the radiation effect. These trapped holes creates conduction, which leads to "ON" state even when $V_{GS}=0V$ [3].

Figure **1**a shows the schematics of a MOSFET floating gate transistor (FGT). The only difference with the standard MOSFET is the addition of a new gate, called the floating gate, between the original gate and the channel. The original gate (topmost) is now called the control gate. A floating gate is basically a polysilicon gate surrounded by insulator and it has no electrical connection with other layers. The working principle of a FGT is almost same as the conventional MOSFET, where the source-drain current is monitored and controlled by the control gate voltage. The floating gate voltage or in other words the stored charge on the floating gate can control the channel between the drain and the source. Thinner tunnel oxide is required to facilitate tunneling between the channel and the floating gate. To program or write a FGT (Figure 1b), a positive gate voltage is applied. This positive voltage attracted electrons from the channel through the tunnel oxide insulating layer into the floating gate. Charge accumulated in the floating gate is protected by the insulating layer. So, the stored data is retained for years. To erase the data, a negative voltage is applied at the control gate (Figure 1c). This negative voltage pushes the electrons out of the floating gate [7].

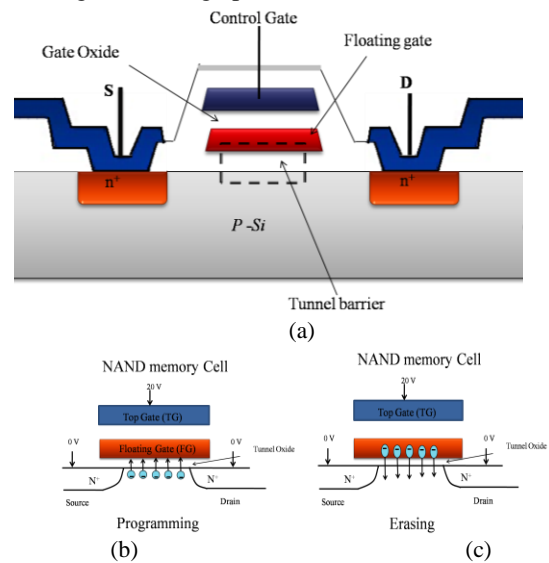

(a)

(b) Programming    (c) Erasing

Figure 1: (a) A floating gate transistor, (b) programming of floating gate transistor, and (c) erasing of floating gate transistor [7].

In this paper, we present how threshold voltage of the FGT is affected by the radiation. To the best of our knowledge, this is the first work to present the analysis of radiation hardness of a flash memory device. An analytical model has been developed, which shows direct relation between radiation level and threshold voltage ($V_{th}$). This equation directly shows how dynamic characteristics are changed due to the radiation exposure. The rest of the paper is organized as follows. Section II presents the effects of radiation on the FGT. Section III provides the results and analysis. Finally, Section IV concludes the paper with a brief introduction to future work.

## II. EFFECTS OF RADIATION ON FGT
### A. Mechanism of Radiation Effect

The changes in characteristics of a FGT due to radiation can be explained by four major steps as shown in Figure 2. The FGT is a modified MOS structure. Therefore, according to device point of view, the radiation effects on the FGT can be explained by the simplified MOS structure.

Step-1: Schematic energy band diagram for MOS structure is illustrated to understand each step of $V_{TH}$ variation. When a positive voltage is applied to the control gate, the electrons flow towards the floating gate and holes to the substrate. Due to irradiation, electron hole pair is generated in $SiO_2$, which is considered as the most sensitive part in the MOS structure. As electrons are more mobile than hole, they swept out in picoseconds or less. In the first picosecond, recombination of electrons and holes takes place; and holes that escape from recombination are relatively immobile and remain near the point of generation, and these holes cause the negative threshold shift in the MOS transistor. This initial stage leads to the maximum drift in the $V_{TH}$ [3].

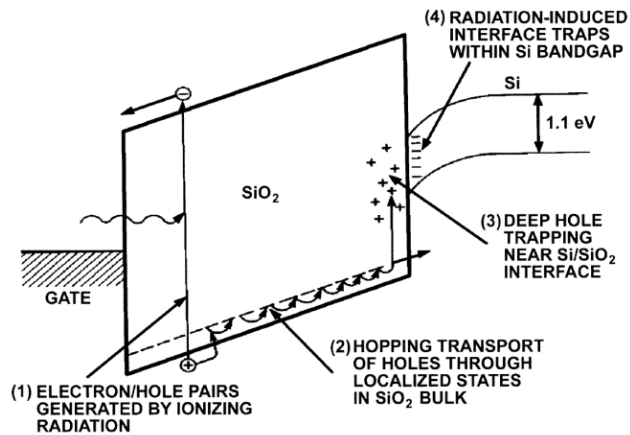

Figure 2: The change of band energy inside the FGT under irradiation [3].

Step-2: Holes tends to shift towards $Si/SiO_2$ interface that causes short-term recovery in the $V_{TH}$, which depends on mainly applied electrical field, temperature and oxide thickness. Generally, it takes about 1sec at room temperature but it may need extended time at low temperature [3].

Step-3: Holes reach to silicon interface, fraction of holes transported to deep long lived trap states. These trapped holes cause the threshold voltage to make a negative shift. This effect continues for hours to years. Gradual annealing can recover the memory from this damage [3].

Step-4: The radiation induced traps at the $Si/SiO_2$ interface are determined by the Fermi level. Generally interface traps are highly dependent on oxide processing [3].

Thus, it can be summarized that radiation induces charge in the oxide, which is dependent on several physical mechanisms like electron-hole pair generation and elelction-hole recombination. The electron-hole recombination depends on the applied electrical field and linear energy transfer.

### B. $V_{TH}$ Variation

Figure 3 shows the distributions of the threshold voltage ($V_{TH}$) for the memory device in both programmed ("0") and erased states ("1"). Here the impacts are shown before and after the exposure to radiation. The $V_{TH}$ of the FGT in the programming ("0") state is high because a large amount of electrons are stored in the floating gate. As a consequence, higher control gate voltage is needed to form the channel in the "0" state. In the programming ("0") state, an electron-hole pair needs 17eV energy around the floating gate (FG) and oxide region by photoemission, which is defined by a process where electrons are emitted from solids under irradiation with photons of sufficiently low wavelength and high energy. Under irradiation, the threshold voltage of programming ("0") state reduces uniformly. Therefore, a comparatively lower control gate voltage creates the conducting channel when the FGT is affected by radiation. On the other hand, the reduction of due to radiation $V_{TH}$ in the erased state ("1") is less prominent.

Portion of the generated electron-hole pairs are recombined, which depends on the electric field around the oxide. Higher electric field leads to lower recombination as more electrons can escape from the recombination. The photoemission is responsible for injection of holes, which escape from the recombination into the FG. These positive holes are recombined partially with the negative electron in the FG. The electrons, which are stored in the FG, get enough energy to jump over the oxide layer barrier when exposed to radiation and by photoemission [1].

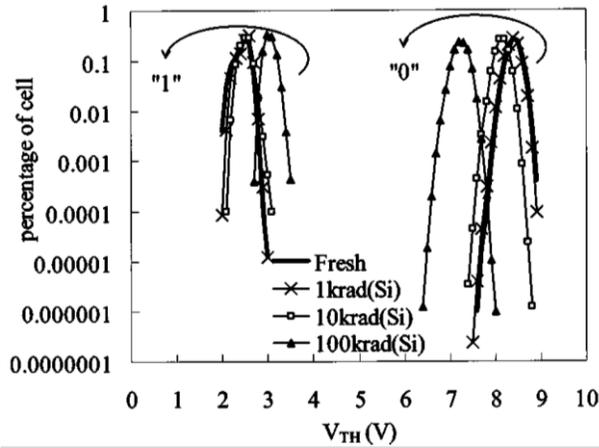

Figure 3: Probability distribution of the threshold voltage for the FGT device for both programming states before and after irradiation [1].

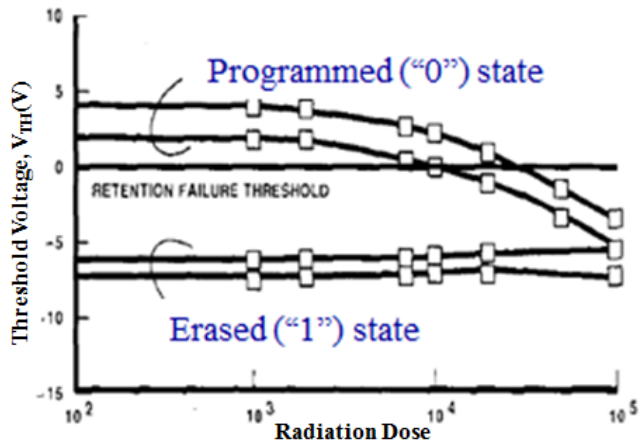

Figure 4 : Threshold voltage data as a function of radiation dose [2].

To observe the $V_{TH}$ variation due to radiation effect, in [2] a memory cell is exposed to cobalt-60 radiation with a dose rate greater than 100rad.s$^{-1}$. The $V_{TH}$ variations are observed over 100krad. The $V_{TH}$ in "0" state is seen to go down significantly even to negative values (see Figure 4). While for the "1" state, the $V_{TH}$ increases slightly. Figure 4 shows that the $V_{TH}$ in programming state ("0") goes down whereas the $V_{TH}$ in the erasing ("1") state increases slightly due to the radiation effect. Therefore, under high radiation doses the logic "0" can be read as logic "1" incorrectly [2].

### C. Time Dependent Effect

Even the radiation effect changes over time, i.e. the change of threshold voltage ($\Delta V_{TH}$) varies over the $10^{-6}$~$10^8$ Sec time scale. Figure 5 shows that $\Delta V_{TH}$ is not fixed after radiation exposure. When $V_{GS}>0V$, holes are trapped into the oxide due to the radiation effect. These trapped holes shifts the operation of the FGT "OFF" to "ON" state even when $V_{GS}=0V$ [3]. Therefore, it gives a wrong reading.

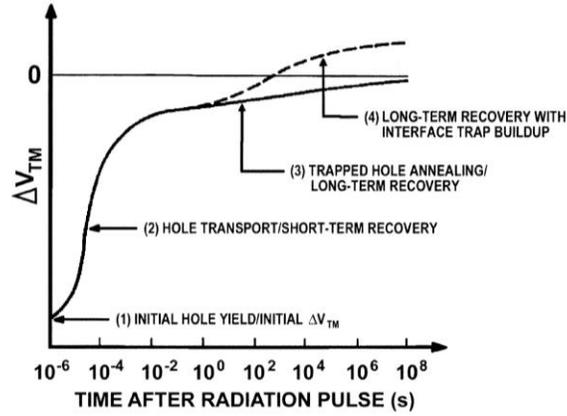

Figure 5: Time dependent post irradiation threshold voltage change of the FGT [3].

### III. RESULT AND ANALYSIS

The change of $V_{TH}$ depends on the charge loss of the FG, which is caused by the photoemission and electron/hole pair generation in the tunnel oxide and control oxide. The change of $V_{TH}$ can be expressed by (1).

$$\Delta V_{TH} = \frac{\Delta Q}{C_{FG}} = \frac{\Delta Q_{TO} + \Delta Q_{CO} + \Delta Q_{PH}}{C_{FG}} \quad (1)$$

Here, $\Delta V_{TH}$ is the change of the threshold voltage, $\Delta Q$ is the total charge loss, $C_{FG}$ is the capacitance between the floating gate and the control gate, $\Delta Q_{TO}$ and $\Delta Q_{CO}$ are the charge losses in the tunnel oxide and control oxide respectively, and $\Delta Q_{PH}$ is the charge loss due to photo emission.

According to the conventional FGT geometry (Figure 1a), the horizontal area of the FG is parallel to the substrate and the lateral area is perpendicular to the substrate. In the existing semiconductor industry, horizontal and vertical FG areas are equal [1]. Radiation causes both electron and hole generation in the surrounding oxides. Therefore, $\Delta Q_{TO}$ and $\Delta Q_{CO}$ linearly depend on both $A_{FGH}$ and $A_{FGV}$. $\Delta V_{TH}$ depends on charge density per area, rather than on the absolute number of stored electrons [9]. In principle, photoemission can happen wherever the electric field is nonzero, i.e., it can depend on both the planar and lateral dimensions of FG. Equation (1) can be rewritten as in (2).

$$\Delta V_{TH} = \frac{(\Delta Q_{TO}).(A_{FGH}) + (\Delta Q_{CO}).(A_{FGH}) + (\Delta Q_{PH})}{C_{FG}} \quad (2)$$

Here, $A_{FGH}$ is the horizontal area and $A_{FGV}$ is the vertical area of the floating gate. Here, $C_{FG}$ is the function of thickness and process of the control oxide. It should be noted that $\Delta V_{TH}$ equation of FGT is completely different from charge trap memory, which explained in [1], because of the structural and charge trapping mechanism differences.

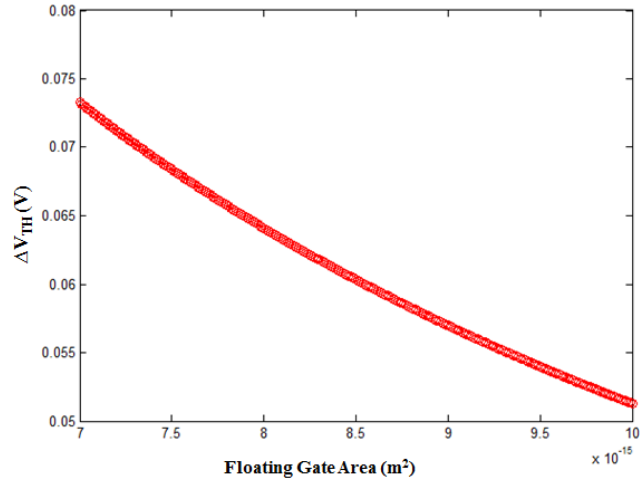

Figure 6: $\Delta V_{TH}$ variation as a function of the floating gate area.

Figure 6 shows $\Delta V_{TH}$ variation with respect to floating gate area for a fixed radiation exposure and oxide thicknesses. It is observed that $\Delta V_{TH}$ is inversely proportional to the floating gate area. 20 nm thick $SiO_2$ control oxide is considered for the computation. The floating gate area, $A_{FG}$ is varied from 0.007~0.01 $\mu m^2$. The above-mentioned values are industry standard, which leads to better $\Delta V_{TH}$ estimation. For convenience, the radiation exposure is assumed fixed and the fringing effects are neglected.

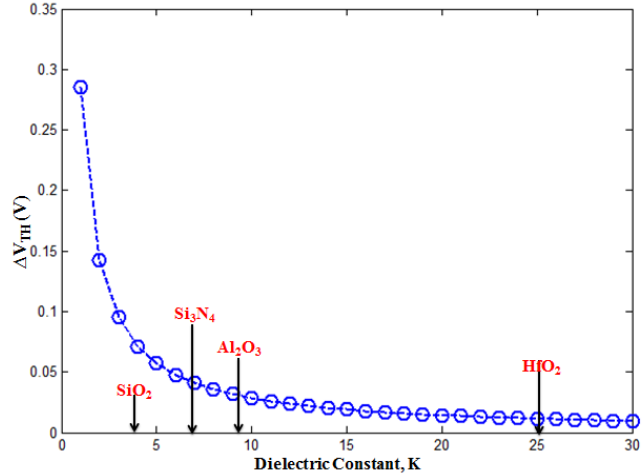

Figure 7: $\Delta V_{TH}$ shift as a function of dielectric constant.

Figure 7 illustrates how threshold voltage changes for different oxide materials. It clearly shows that high-K oxides exhibits low $V_{TH}$ shift which leads high radiation hardness. Therefore, high-k oxide is recommended as the control oxide to make the circuit radiation hard. A comparative study is done for the most popular oxides [6] ($SiO_2$, $Si_3N_4$, $Al_2O_3$ and $HfO_2$) in the current semiconductor industry. This study suggests that if high k-dielectric oxide is used as the control oxide, the $V_{TH}$ variation tends to be less and at a certain higher value of dielectric constant (k) the variation tends become zero, which leads to better radiation hardness. According to the analysis, $HfO_2$ is the best control oxide choice for flash memory when radiation hardness is the major concern.

Many researches have shown that how $V_{th}$ changes after the device is kept under radiation. For CAD tool and IC designer community it is required to translate the radiation effect quantitatively. Keeping that in mind, we have considered a black box where a FGT/MOSFET is kept as shown in **Figure 8**. The role of the model is to compute $V_{TH}$ values for increasing radiation levels for given device which is already fabricated or designed i.e. other parameters will not change. We are stable to the condition because the experimental results which are available followed the approach.

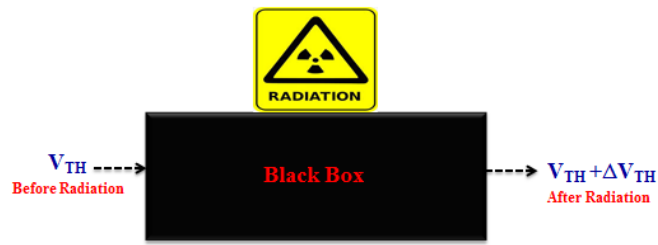

Figure 8: Black box of a flash memory under radiation exposure.

The variation of the $V_{TH}$ as a function of radiation data are collected from the experimental results [2],[4]. Then the data is statistically analyzed. These steps and data are not provided in the paper because of space limitation. The statistical analysis is concluded by the result shown Figure 9, where x-axis presents TID (dose of radiation in the $Krad(SiO_2)$ unit) and y-axis represents $V_{TH}$.

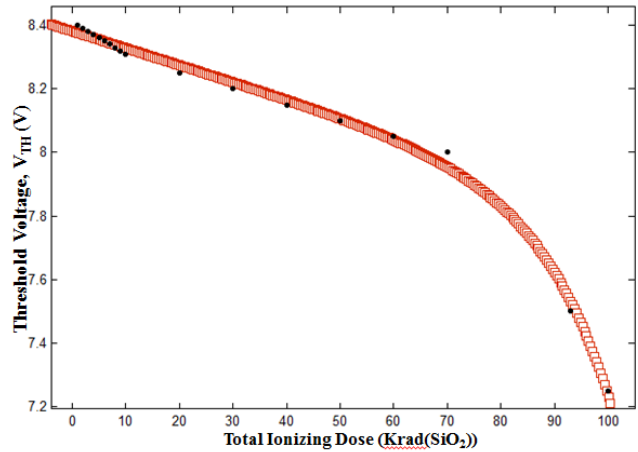

Figure 9: Threshold shift as a function of total ionization dose (TID).

Figure 9 shows that $V_{TH}$ decreases as a function of the radiation level. As the radiation exposure increases, $V_{TH}$ tends to fall rapidly. The black dots are the experimental value collected from [4], while the continuous red curve represents the simulated result. It should be noted that TID(Krad(SiO$_2$)) is a well-defined universal radiation measurement unit, which is very popular in experimental and commercial radiation measurement. In order to validate the model, the result is verified with experimental research works. The simulated result of the $V_{TH}$ variation with radiation shows good agreement with the experimental data of [2],[4].

## IV. CONCLUSION AND FUTURE WORK

A mathematical model of FGT is proposed where the threshold voltage ($V_{TH}$) is considered as the key parameter. The $V_{TH}$ of the FGT drops when radiation exposure rises. From our analysis, we observed that the variation of $V_{TH}$ in the FGT is (i) inversely proportional to the floating gate area, (ii) directly proportional to the control oxide thickness, and (iii) drops exponentially at the higher value of dielectric constant. Therefore, the mathematical model will be useful to analyze the radiation hardness of flash memory design and allow trade-off between important parameters. Our future work involves the radiation hardness test at every single design step of a device which will allow designers more flexibility in the radiation hardened memory design in future.